\documentclass[11pt]{article}
\usepackage{graphicx}

\def\be{\begin{equation}}
\def\ee{\end{equation}}
\def\ba{\begin{eqnarray}}
\def\ea{\end{eqnarray}}

\def\12{{1\over 2}}

\def\msun{M_\odot}

\def\ltsima{$\; \buildrel < \over \sim \;$}
\def\simlt{\lower.5ex\hbox{\ltsima}}
\def\gtsima{$\; \buildrel > \over \sim \;$}
\def\simgt{\lower.5ex\hbox{\gtsima}}

\begin{document}

\title{\bf Observational Manifestations of the First Protogalaxies in the 21 cm Line\footnote{
Astronomy Reports, 2012, Vol. 56, No. 2, pp. 77-83.}}
\author{E.~O.~Vasiliev$^{1,2}$\thanks{eugstar@mail.ru}, Yu.~A.~Shchekinov$^{2,3}$ \\
\it $^1$Institute of Physics, Southern Federal University, Rostov-on-Don, Russia \\
\it $^2$Physics Department, Southern Federal University, Rostov-on-Don, Russia \\
\it $^3$Special Astrophysical Observatory, Russian Academy of Sciences, \\
Nizhnyi Arkhyz, Karachay-Cherkessia Republic, Russia}

\date{}

\maketitle

\begin{abstract}
The absorption properties of the first low-mass protogalaxies (mini-halos) forming at high
redshifts in the 21-cm line of atomic hydrogen are considered. The absorption properties of these
protogalaxies are shown to depend strongly on both their mass and evolutionary status. The optical depths in the line
reach $\sim$0.1-0.2 for small impact parameters of the line of sight. When a protogalaxy being compressed, the
influence of gas accretion can be seen manifested in a non-monotonic frequency dependence of the optical
depth. The absorption characteristics in the 21-cm line are determined by the thermal and dynamical
evolution of the gas in protogalaxies. Since the theoretical line width in the observer's reference frame is
1-6 kHz and the expected separation between lines 8.4 kHz, the lines from low mass protogalaxies can
be resolved using ongoing and future low frequency interferometers.
\end{abstract}



\section{Introduction}

\noindent

One of the means of studying the history of intergalactic
space before the epoch of secondary ionization
(reionization) of hydrogen is radio observations
in the 21-cm line of neutral hydrogen [1-3]. At
least three methods for studying this epoch have been
proposed: (1) observations in emission or absorption
in the neutral intergalactic medium against the global
cosmic microwave background (CMB) signal [4, 5],
(2) statistical studies of the angular distribution of
the signal fluctuations [6, 7], and (3) observations
of absorption in the spectrum of a distant strong
radio source, which may reveal a ``forest'' in the 21-
cm line (analogous to the L$\alpha$ forest in the optical
spectra of quasars and galaxies) [8-10]. The first
two methods involve mapping spatial (position on the
sky) and temporal (redshift) properties of the intergalactic
medium, while the latter provides information
only along a line of sight as a function of redshift.
Since the spatial scales corresponding to the angular
resolutions of ongoing or futture radio telescopes, such as LOFAR\footnote{http://www.lofar.org/index.htm} are roughly
1~Mpc, only large-scale structures, such as giant ionization
regions around massive galaxies and quasars,
can be observed. Measurements of absorption at
different redshifts can be used to resolve small-scale
structures, individual protogalaxies (mini-halos), and
ionization zones in low-mass protogalaxies.

The effect of external ionizing radiation on the
properties of the 21-cm forest was studied in [9-14]. Mainly, the dark-matter profiles in protogalaxies
have been taken to be steep (cusped) [15]. However,
observations of dwarf galaxies in the local Universe
indicate a ``flat'' distribution of dark matter (a truncated
isothermal sphere with a flat core) [16]. Recent
theoretical studies have also found that the first protogalaxies
could have flat dark-matter profiles [17-19].
The first halos were assumed in [9, 10] to be virialized
and isothermal, and their absorption characteristics
were calculated without accounting of the gas and dark matter evolution. 
Some evolutionary effects
were included in [14], such as a possible influence of
gas accretion and star formation, although the 
gas and dark matter profiles of minihalos before 
the first stars have appeared were assumed to
be time-independent. It is well known though that the density
inside the perturbation from which a protogalaxy does 
form increases by more than a factor of 200 in the
non-linear stage, facilitating cooling of the gas and
a decrease of its ionization fraction [20-23]. It is
obvious that such appreciable evolutionary variations
in the physical properties of the gas could substantially affect 
absorption properties of the first protogalaxies. With this in mind, Meiksin [13]
considered possible variations in the absorption properties
caused by the virialization process of the first protogalaxies.

Here, we investigate the absorption properties of
the first protogalaxies within a self-consistent, one-dimensional,
spherically symmetric description of their thermal and chemical evolution.
In the heirarchical scenario of galaxy formation, the
first protogalaxies able collapse and lead to  
formation of the first stars had low masses. The
minimum mass of the first protogalaxies,  $M_h\sim 10^7~M_\odot$ 
at redshifts $z\sim 10$, was apparently determined
by molecular-hydrogen cooling [23]. We will study
the 21-cm profiles for mini-halos with masses near
this minimum mass, with the aim of using possible
differences in the 21-cm absorption characteristics
as a tool to answer the question of whether the minimum
masses of halos in which star formation began
were indeed close to $M_h\sim 10^7~M_\odot$  at $z\sim 10$. Section 2
describes the model used, Sections 3 and 4 present
our results and conclusions.
The computations assume a $\Lambda$CDM cosmological model with 
 $(\Omega_0,\Omega_{\Lambda},\Omega_m,\Omega_b,h ) 
= (1.0,\ 0.76,\ 0.24,\ 0.041,\ 0.73 )$, and
a relative number density of deuterium of 
$n[{\rm D}]/n = 2.78\cdot 10^{-5}$ [16].


\section{Evolution of minihalos}

\noindent

We modeled evolution of the first protogalaxies
using a one-dimensional Lagrangian approach similar
to that proposed in [23]. It is difficult to fully take
into account mergers of protogalaxies and other essentially 
non-one-dimensional dynamical effects in
such a simple model. This would require self-consistent,
three-dimensional cosmological simulations with
high resolution, which is not currently possible. By necessity 
therefore we will use simplified model approximations that
are commonly applied to reduce multi-dimensional
problems to spherically symmetric ones.

\subsection{Dynamics of Dark Matter}

The description of dark matter in our model is
analogous to that of [23]. It is assumed that dark matter
with the mass $M_{DM} = \Omega_{DM} M_{halo}/\Omega_m$ is contained inside
the radius $R_{tr}$ and has a truncated 
isothermal sphere profile with a flat core of radius $R_{core}$, 
here $\Omega_{m}$ is the matter density and $\Omega_{DM} = \Omega_{m} - \Omega_{b}$. 
The ratio $\xi = R_{core} /R_{vir}$ is assumed through the paper to be 0.1; 
$R_{vir}$ is the
virial radius of the halo, outside the radius $R_{tr}$ ,
the dark-matter density is equal to the background
cosmological value, $\rho_{DM} = \rho_0 \Omega_{DM} (1+z)^3$, 
where $\rho_0 = 1.88\times 10^{-29} h^2$ is the critical density; 
the radii $R_{tr}$ and $R_{core}$ evolve in time [23]. Such a
description is usually accepted to imitate the evolution of
density perturbations (see, e.g., [25]). Before the ``turnaround'' point, 
i.e. the point where the perturbation separates from the
Hubble flow, the density within the truncated sphere,  
$R<R_{tr}$, varies with time similarly to what approximated in [21] 

\be
\rho = \rho_0 \Omega_{dm} (1+z)^3 {\rm exp}\left({2.8 A \over 1 - 0.5A^2}\right), 
\label{rhot97}
\ee
where $A = (1+z_{vir})/(1+z)$. We accept in (\ref{rhot97}) the numerical 
coefficients in the exponent slightly different from the ones used in [21], 
in order to achive a better agreement between (\ref{rhot97}) and 
the exact solution in the vicinity of the turnaround
point. The approximation (1) violates 
near the virialization point, however this is not important
for our purposes, since it is not used beyond the turnaround point.

\subsection{Gas Dynamics and Chemical Kinetics}

For modeling evolution of the gas (i.e., the baryonic component of 
a protogalaxy) a one-dimensional spherically symmetric 
Lagrangian scheme similar to that described in [26] is used.
The standard resolution in the computations is 
1000 cells (spherical shells), which provides a sufficiently
good convergence of the scheme.

Chemical kinetics of the primordial gas includes
the following main components H, H$^+$, $e$, H$^-$, He, He$^+$, He$^{++}$, 
H$_2$, H$_2^+$, D, D$^+$, D$^-$, HD, HD$^+$. The
corresponding reaction rates are taken from [27, 28].
In the energy equation radiative
losses typical for the primordial gas: cooling
with recombination, collisional excitation of hydrogen
and helium, free-free transitions, Compton interactions
with CMB photons [29], and molecular cooling
via H$_2$ [28] and HD [30, 31] are accounted.

All models begin at the redshift $z = 100$,
when the density perturbations are small and obviously
linear. The initial parameters for $z = 100$, i.e.,
the gas temperature, ion and molecular composition,
etc, are found from a one-zone calculateion starting
at $z = 1000$ with the standard gas parameters for the
end of the recombination epoch: $T_{gas} = T_{CMB}$, 
$x[{\rm H}] = 0.9328, x[{\rm H^+}] = 0.0672, x[{\rm D}] 
= 2.3\times 10^{-5}, 
x[{\rm D^+}] = 1.68\times 10^{-6}$ 
(see [23, Table 2] for details
and references). The initial size of the computational
domain is 10 virial radii for a protogalaxy of a given 
mass.

\subsection{The 21 cm Line}

Only collisional excitation by atoms and electrons
is taken into account when computing the spin
temperature for the 21-cm line [32]:

\be
T_s = {T_{CMB} + y_c T_k \over 1 + y_c} 
\ee
where $T_{CMB}$ is the temperature of the CMB, $y_c$ is
a function determined by the efficiency of collisional
excitation 

\be
y_c = {C_{10}T_\ast \over A_{10} T_k}, 
\ee
$T_\ast = 0.0682$~K is the energy for the transition between
sublevels of the hyperfine structure, $A_{10} = 2.87\times 
10^{-15}$~s$^{-1}$ is the transition probability, 
$C_{10} = k_{10}n_H + \gamma_e n_e$ is the rate of collisional de-excitation
by hydrogen atoms and electrons (we neglect the
contribution from protons), the approximations 
for $k_{10}$ of [33] and for $\gamma_e$
from [34] are used. The computations do not take into account
the influence of resonance ultraviolet radiation (UV
pumping) on the level populations [35], since at redshifts $z = 10-15$ this
effect can be important only locally near
galaxies with star formation. At lower redshidt the regions of penetration of UV radiation from
the stellar population become substantially greater and
overlap and manifest reionization of
the Universe. Moreover, in order for the contribution
from UV pumping to the variations of the spin temperature
to be comparable to the contribution from
collisions, the flux of resonance UV radiation must be
fairly high: $F_{Ly\alpha} \simgt 0.1$~erg~s$^{-1}$~cm$^{-2}$ (this estimate is
for typical temperatures $T\sim 200$~K and gas densities
$n\sim 1$~cm$^{-3}$ in protogalaxies at $z \sim 10$).


\section{Absorption in the 21 cm line}

\noindent

We will study two observable quantities that describe
the absorption properties of the gas in protogalaxies:
the optical depth at frequency $\nu$ along
the line of sight, and the total absorption within the
protogalaxy, i.e., the line equivalent width. We
consider the properties of low-mass protogalaxies
(mini-halos) with masses of $10^5$, $10^6$ and $10^7\msun$,
which separated from the Hubble flow at redshift
$z_{ta} \simeq 15.5$ and virialized by $z_{v} \simeq 10$.

\begin{figure}
\center
\includegraphics[width=15cm]{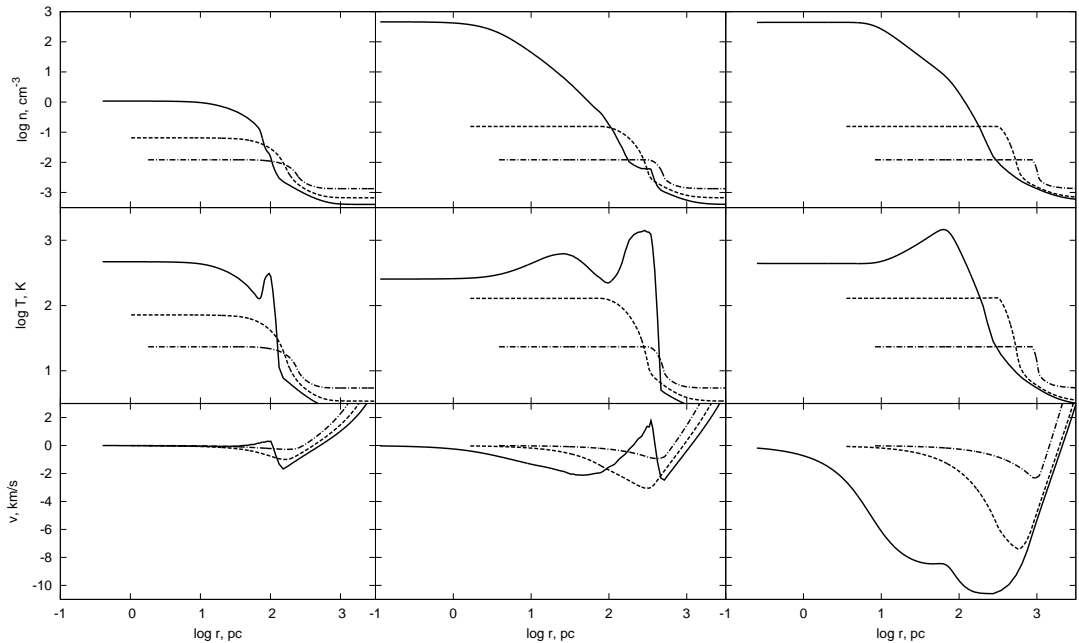} 
\caption{
Number density, temperature, and velocity (from top to bottom) of the gas in protogalaxies with masses of $10^5$, $10^6$ and $10^7\msun$ (from left to right) at times $z_{ta} = 15.5$ (dot-dashed curve), $z = 12$ (dashed curve), and $z_v = 10$ (solid curve).
}
\label{figevol} 
\end{figure}

Evolution of the first protogalaxies depends
critically on their masses: low-mass protogalaxies
cannot cool rapidly and form stars after their virialization.
With increasing protogalaxy mass the
number density of hydrogen molecules in the central
regions grows, facilitating cooling of the gas and
formation of the first stars (see [20-23, 36] for
more detail). Figure 1 presents radial profiles of the
gas density, temperature, and velocity 
in protogalaxies with masses of $10^5$, $10^6$ and $10^7\msun$
at times close to the turnaround ($z_{ta} = 15.5$)
and the virialization ($z_{v} = 10$), and also at a time
corresponding to the intermediate redshift ($z = 12$).
When virialized at $z_{v} = 10$, only protogalaxies with
masses $M\simgt 10^7\msun$ can form stars [22, 23]. In the
process of virialization, gas density in the
central regions already reaches $n\simgt 10^8$~см$^{-3}$ at $z \simeq 11$, 
which essentially corresponds to the collapse and the
onset of formation of the first pre-stellar cores.
Therefore, for the virial state of the mass
$M= 10^7\msun$ we present the absorption characteristics
corresponding to $z \simeq 11$, while for the two other masses the 
characteristics are calculated for $z_v = 10$. As 
noted above, in lower-mass protogalaxies, the prestellar
cores either form after virialization or do not
form at all.

The optical depth at frequency $\nu$ along a line of
sight passing a distance $\alpha$ from the center of the
protogalaxy is
\be
 \tau_\nu = {3h_p c^3 A_{10}\over 32 \pi k \nu_0^2} \int_{-\infty}^{\infty}
 {dR {n_{HI}(r) \over \sqrt{\pi} b^2(r) T_s(r)} 
    {\rm exp}\left[{-{[v(\nu) - v_l(\alpha,R)]^2\over b^2(r)}}\right]}
\label{optd}
\ee
where $r^2 = \alpha^2 + R^2$, $v(\nu)=c(\nu-\nu_0)/\nu_0$, $v_l(\alpha,R)$ is
the gas infall velocity projected on to the line
of sight, $b^2 = 2kT_k(r)/m_p$ is the Doppler parameter,
and the remaining notations are standard.

\begin{figure}
\center
\includegraphics[width=15cm]{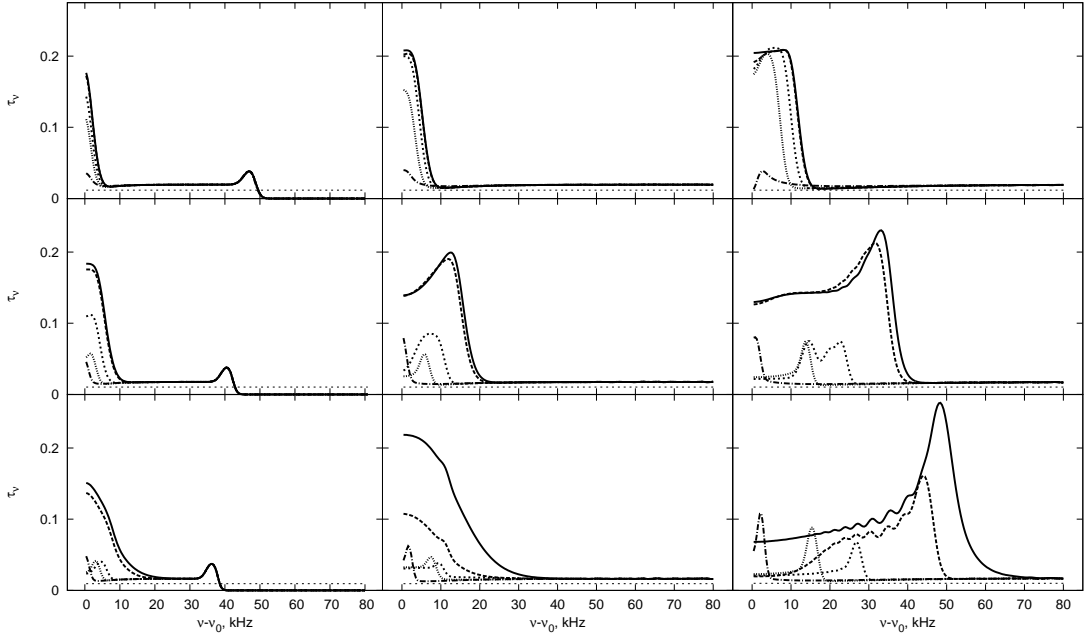} 
\caption{
Optical depth for protogalaxies with masses of $10^5$, $10^6$ and $10^7\msun$ 
(from left to right) at times $z_{ta} = 15.5$ (upper panel), $z = 12$ (middle panel), 
and $z_v = 10$ (lower panel). The curves correspond to the line of sight impact 
parameters  $\alpha = 0.1, 0.3, 1, 1.5, 3r_{vir}$  (from right to left), and 
the thin dashed line to the background optical depth.
}
\label{figoptd} 
\end{figure}

Figure 2 shows the optical depths for lines of
sight passing at several impact parameters for protogalaxies
with masses of $10^5$, $10^6$ and $10^7\msun$ at
times close to the turnaround ($z_{ta} = 15.5$) and
the virialization time ($z_v = 10$), as well as at a time
corresponding to the intermediate redshift $z = 12$.
Considerable differences in the line absorptions 
are clearly visible, both for protogalaxies with different
masses and during the evolution of a protogalaxy of
a given mass. At $z = 15.5$ (upper panel 
in Fig. 2), the gas densities and temperatures 
in the protogalaxies with different masses are
nearly the same (Fig. 1), and the differences in the line
profiles are determined by the distributions of the gas velocity:
with increasing protogalaxy mass the accretion velocity 
increases from $\sim$0.2 km/s for $10^5\msun$ to
$\sim$2 km/s for $10^7\msun$. The small peak
in the optical depth near $\nu-\nu_0 \simeq 40$~kHz and the
drop to zero at higher frequencies for $M = 10^5\msun$ is, from one side, 
due to the drop in the temperature at the periphery
of the protogalaxy and the transition to the background
value [a decrease in the denominator in the
exponent (4)],  and from the other due to the transition of the velocity profile to
the Hubble expansion [an increase in the
numerator in the exponent (4)]. In more massive
protogalaxies, a similar peak in optical depth is
located at higher frequencies.

The absorption characteristics change considerably as
the protogalaxy is compressed. At redshift $z = 12$
(middle panel in Fig. 2), the absorption lines
become broader, and features due to the gas dynamics 
during the collapse of the protogalaxies appear.
In low-mass protogalaxies ($M = 10^5\msun$), the enhancement
of gas velocity in the accretion shock is modest, so that the absorption line
profiles for the various impact parameters of the
line of sight broaden only slightly. In higher-mass
protogalaxies ($M \simgt 10^6\msun$), the accretion wave is
stronger, and a stronger jump in the temperature and
density arises, which manifests a nonmonotonic
frequency dependence of the optical depth for the lines
of sight intersecting this region ($\alpha \simlt r_{vir}$) as,
more precisely, a decrease in the optical depth in the
line center.

At $z = 10$ (at $z \simeq 11$ for $M = 10^7\msun$), the
absorption-line profiles depend substantially on differences
in the dynamical and thermal evolution
of the protogalaxies (lower panels in Figs. 1
and 2): the lines become appreciably broader, the
strongest absorption is observed for $\alpha \simlt 0.3r_{vir}$,  
the influence of gas accretion is clearly seen in the
protogalaxy with mass $M = 10^7\msun$ resulting in pronounced peaks
in the frequency dependence of the optical
depth. In more massive protogalaxies gas cools more efficiently, 
and, moreover, the shocks associated with virialization appear to be
stronger in more massive protogalaxies. Note that the
line profile for $M = 10^6\msun$ at $z = 10$
has changed since $z = 12$: contrary to $z=12$ at $z=10$ the optical
depth falls off essentially monotonically from the line
center. This is due to a decrease in the accretion, and 
the appearance of a reverse shock acting to oppose
the accretion (Fig. 1). In the case of higher-mass
protogalaxies ($M = 10^7\msun$), the accretion increases
with decreasing redshift, while the non-monotonic
dependence of the optical depth becomes stronger.

\begin{figure}
\center
\includegraphics[width=15cm]{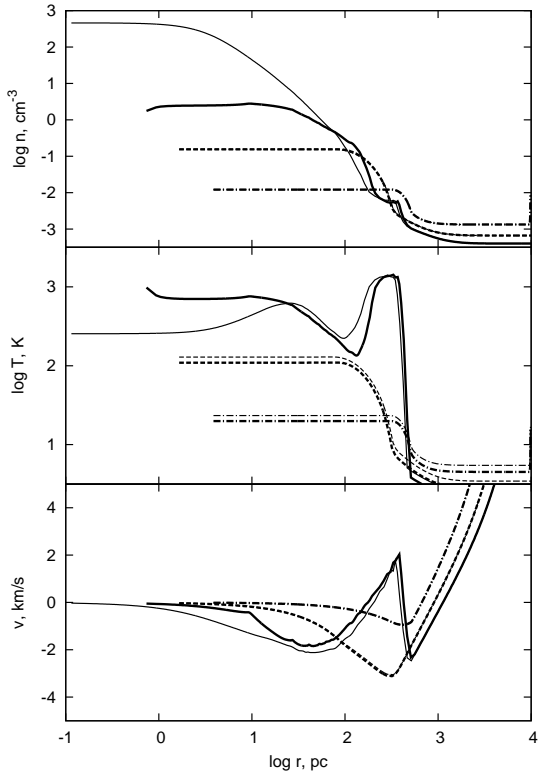} 
\caption{
Density, temperature, and velocity (from
top to bottom) in a protogalaxy with mass $10^6\msun$ at
times $z_{ta} = 15.5$ (dot-dashed curve), $z = 12$ (dashed
curve), and $z_v = 10$ (solid curve), with (thin curves) and
without (thick curves) taking into account cooling of the
gas. The difference is clearly seen at the virialization
time $z_v = 10$; these curves coincide at the other two times
considered.
}
\label{figevol16} 
\end{figure}

Let us now investigate how thermal evolution
affects the absorption-line profiles. For this purpose, we
calculate the evolution of a protogalaxy with mass
$M = 10^6\msun$ without radiation gas cooling in 
atomic and molecular processes (i.e. we consuder the adiabatic
case), and compare it with the results
described above (Fig. 3). In the presence of radiation cooling the central
regions of the protogalaxy are seen to cool efficiently during virialization,
and the radial dependences differ considerably 
from the adiabatic case, but these differences are
substantial only in the inner region of the protogalaxy,
$\alpha \simlt 0.5r_{vir}$, where the effects of cooling are stronger
due to the higher density. As a result, the absorption line
profiles will also differ in this region as 
seen in Fig. 4.

\begin{figure}
\center
\includegraphics[width=15cm]{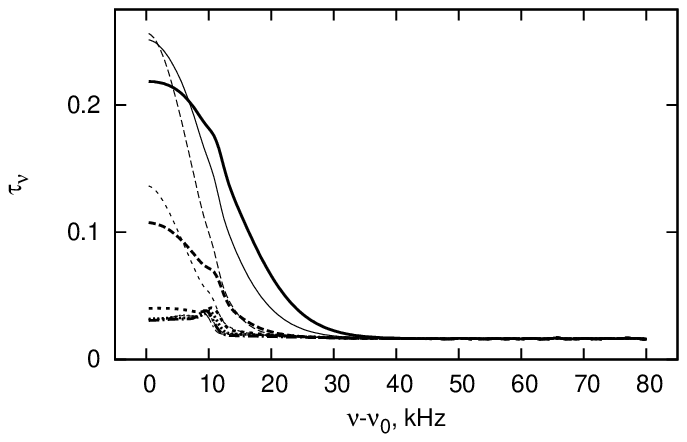} 
\caption{
Optical depth for a protogalaxy with
mass $10^6\msun$ at time $z_v = 10$. The curves
correspond to the line of sight impact parameters 
$\alpha = 0.1, 0.3, 0.5, 0.8, 1 r_{vir}$  (from
right to left) with (thin curves) and without (thick curves)
radiation gas cooling. The difference is
clearly seen for $\alpha \simlt 0.5 r_{vir}$; the curves coincide for
higher values of $\alpha$.
}
\label{figoptd16} 
\end{figure}

Another observational characteristic is the line
equivalent width, $\langle\Delta\nu\rangle_{obs} 
= \langle\Delta\nu\rangle_i / (1+z)$, where
the rest-frame equivalent width is
\be
 {\langle\Delta \nu_i\rangle \over 2} = \int_{-\infty}^{\infty}{(1-e^{-\tau_\nu})d\nu} 
 - \int_{-\infty}^{\infty}{(1-e^{-\tau_{IGM}})d\nu} 
\ee
where $\tau_{IGM}$ is the background optical depth of the
neutral intergalactic medium. Figure 5 presents the
observed line equivalent width for absorption in protogalaxies
with masses of $10^5$, $10^6$ and $10^7\msun$ at redshifts
$z = 15.5, 12$, and 10. The equivalent width depends
only weakly on the impact parameter of the line
of sight at $z = 15.5$, but in the process of formation
of a protogalaxy it grows at $\alpha \simlt r_{vir}$ , reaching 0.1-0.4~kHz
for $M = 10^6-10^7\msun$. The line width of the perturbations
after separation from the Hubble flow in the
observer's rest frame (i.e. at $z = 0$) equals to 1-6~kHz.
Since the distance between the lines is $\sim$8~kHz at
$z = 10$ [14], and the spectral resolution of 0.76~kHz can
be reached with existing and planned low-frequency
interferometers, such as LOFAR and the SKA\footnote{http://www.skatelescope.org/} the
lines of low-mass protogalaxies might be quite resolvable 
with these instruments.

\begin{figure}
\center
\includegraphics[width=15cm]{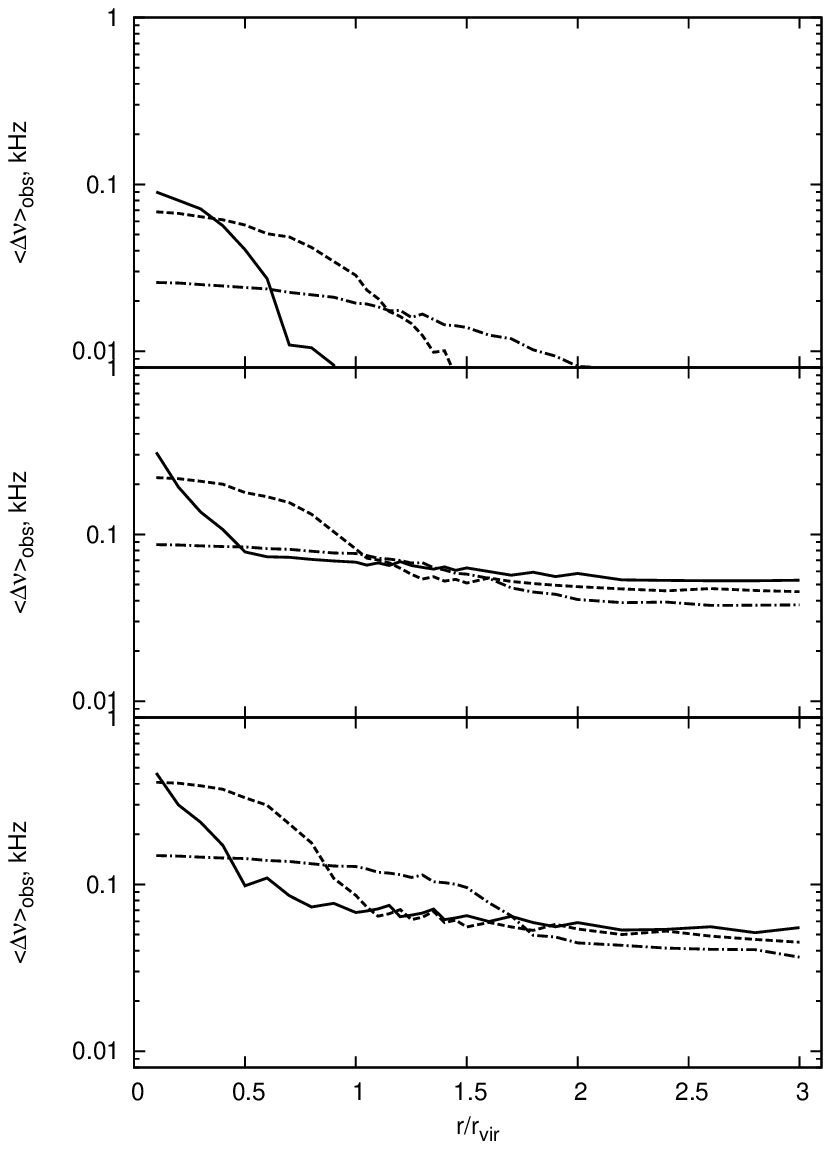} 
\caption{
Observed line equivalent width for protogalaxies
with masses of $10^5$, $10^6$ and $10^7\msun$ (from top to bottom)
at times $z_{ta} = 15.5$ (dot-dashed curve), $z = 12$ (dashed
curve), and $z_v = 10$ (solid curve).
}
\label{figew} 
\end{figure}

Note that when describing the initial growth
of a density perturbation after recombination, it
is assumed that the velocities of the dark matter
and baryons are similar and their relative velocities
low [37]. This picture has been recently supplemented
with a possibility of supersonic relative motions
of baryons and dark matter with velocities $v_s(z) = v_{s,i}/(1+z)$, 
where $v_{s,i}$ is the rms relative velocity
of the gas at the epoch of recombination [38]. It
is assumed that, during the compression of 
protogalaxies, their gas moves with an effective speed
$v_{eff} = \sqrt{c_s^2 + v_s^2}$ ($c_s$ is the sound speed) [39]. We
carried out computations of the evolution of protogalaxies
with masses of $M=10^5 - 10^7~\msun$  taking
into account $v_s(z)$, adopting the characteristic value
$v_{s,i} = 30$~km/s for the epoch of recombination [38].
It was found that compared to the standard case
described above (Fig. 2) the accounting of the relative motion of baryons and dark matter 
results in an approximately
10-30\% decrease in the optical depth in the 21-cm
line at the epoch of separation of the perturbation 
from the Hubble flow. For
more massive protogalaxies this change is smaller and decreases by the epoch of virialization,
while  for low-mass protogalaxies remains considerable comprising about 10\%.

\section{Conclusions}

\noindent

We have used one-dimensional, spherically symmetric 
computations to investigate the absorption
properties of the first protogalaxies in the 21-cm line of HI 
at the lower mass boundary admitting
the collapse of baryons and star formation,
$M_h\leq 10^7~M_\odot$ at $z\simgt 10$. We have studied the evolution
of the 21-cm line profile and observed equivalent
width during the virialization of these protogalaxies.

The absorption properties of the protogalaxies
depend substantially on both their mass and their
evolutionary status. The optical depth in the line
reaches $\sim 0.1 -0.2$ at small line of sight impact parameters 
for protogalaxies with masses of 
$M = 10^5 - 10^6\msun$, then monotonically falls off to the
background value. The 21-cm absorption profiles
for halos with masses of $M=10^7~M_\odot$ and probably
higher masses differ qualitatively from the profiles
of the halos with masses below this lower limit: massive halos show 
a dip at the line 
center. This can be used to observationally determine
the minimum masses of protogalaxies; i.e., of minihalos
capable to cool radiatively, and enabling the
collapse of baryons with subsequent star formation.

The influence of gas accretion, manifested as a nonmonotonic
frequency dependence of the optical depth, can be seen as the
protogalaxies are compressed. 
The accretion wave is observed at large line-of-sight
impact parameters for protogalaxies with masses
$M \simeq 10^5\msun$ , and becomes stronger for protogalaxies
with higher masses ($M \simgt 10^6\msun$), leading to a large
jump in the gas temperature and density, which gives
rise to appreciable peaks in the frequency dependence
of the optical depth for $\alpha \simlt r_{vir}$. The absorption properties
are mostly determined by the thermal and dynamical
evolution of the gas in the protogalaxies.

The equivalent width corresponding to the turnaround point 
in the observer’s rest frame (i.e., $z = 0$) depends
only weakly on the line-of-sight impact parameter $\alpha$, but grows
with the formation (virialization) of protogalaxies with
masses of $M = 10^6-10^7\msun$, reaching 0.1-0.4~kHz
for $\alpha \simlt 0.5 r_{vir}$. The expected distance between
lines is 8.4 kHz at $z\sim 10$ [14], that is larger than
the width of the 21 cm absorption lines from low mass
protogalaxies in the observer’s rest frame, so
that the lines can be resolved using low frequency
ongoing and future interferometers.

\section{Acknowledgements}

\noindent

This work was supported by the Russian Foundation for Basic Research (project codes 09-02-00933,
11-02-90701) and the Ministry of Education and Science of the Russian Federation, in the framework
of the Federal Targeted Program “The Scientific and Science Education Staff of Innovative Russia”
for 2009-2013 (state contracts 02.740.11.0247 and P-685) and the Analytical Departmental Targeted
Program “The Development of the Scientific Potential of Higher Education” RNP-2.1.1/11879.



\end{document}